\documentclass[12pt,preprint]{aastex}

\def\ltsima{$\; \buildrel < \over \sim \;$}
\def\lsim{\lower.5ex\hbox{\ltsima}}
\def\gtsima{$\; \buildrel > \over \sim \;$}
\def\gsim{\lower.5ex\hbox{\gtsima}}

\begin{document}

\title{Transits of Transparent Planets - Atmospheric Lensing Effects}
\author {Omer Sidis$^1$}
\author{Re'em Sari$^{1,2}$}
\affil{$^1$Racah Institute of Physics, Hebrew University, Jerusalem 91904, Israel \\
$^2$Theoretical Astrophysics, Caltech 350-17, Pasadena, CA 91125, USA}

\begin{abstract}
Light refracted by the planet's atmosphere is usually ignored in analysis of planetary transits.
Here we show that refraction can add shoulders to the transit light curve, i.e., an increase in the observed flux, mostly just before and
after transit. During transit, light may be refracted away from the observer. Therefore, even completely
transparent planets will display a very similar signal to that of a standard transit, i.e., that  of an opaque planet.
We provide analytical expression for the amount of additional light deflected towards the observer before the transit,
and show that the effect may be as large as $10^{-4}$ of the stellar light and therefore measurable by current instruments.
By observing this effect we can directly measure the scale height of the planet's atmosphere.
We also consider the attenuation of starlight in the planetary atmosphere due to Rayleigh scattering
and discuss the conditions under which the atmospheric lensing effect is most prominent.
We show that, for planets on orbital periods larger than about 70 days,
the size of the transit is determined by refraction effects, and not by absorption within the planet.
\end{abstract}

\section{Introduction}
While the majority of extrasolar planets were found using the radial velocity technique \citep{MUN+04,MBV+08},
the method of finding planets by observing their transit in front of a star \citep{BuS84} is now becoming a powerful tool.
The transit method is based on the observation of a small drop in the brightness of a star, that occurs when the orbit of a planet passes ('transits') in front of it. The amount of light lost depends on the ratio of the radius of the planet to the radius of the star. It is of order 0.01\% (for earth-size planets) and 1\% (for Jupiter-size planets). Transiting extrasolar planets reveal more information. Since the star's size  can be estimated from spectroscopic observations, the planet's size can be determined.
The main effect  of planetary transit can be modeled geometrically \citep[e.g.,][]{Sea08}. Both star and planets are projected on the plane of the sky, producing two circles of radius $R_*$ and $R_p$. Then, the fractional overlap between the two
determines the fractional decrease in stellar light. Within transit, the size of the effect is $(R_p/R_*)^2$.
The high accuracy obtained, especially by space instruments \citep{BCG+01,BKJ+09} demand more advanced models, which take into account the non uniform stellar brightness over the stellar disk. This results in a curved, rather than a flat, lightcurve around the eclipse minimum. In addition, reflection and thermal emission from the planet \citep{LoS07}
cause an out of eclipse sinusoidal variation in the flux \citep{BKJ+09,KCC+09}. \cite{BHB07} stress that care should be taken in the definition of the planet radius. While the canonical definition uses radial optical depth of $\tau=2/3$,
the light in a transit is absorbed tangentially, and therefore passes a longer path through the atmosphere of the planet than a radial ray would if it were to penetrate to the same depth. This results in an increase of the apparent radius of the planet by about 5 planetary scale heights.

\cite{SeS00} showed that the lengthening of the light path due to refraction in the atmosphere is small and
therefore does not change the absorption significantly.
\cite{HuS02} considered refraction by planetary atmospheres, taking the effects of oblateness and absorption into account.
In this paper we consider the effects of refraction in the planetary atmosphere independent of absorption.
Outside of the transit, refraction may redirect light through the planetary atmosphere towards the observer.
This results in increased observed brightness especially right before and after a transit. In contrast,
during transit, refraction deflects light that could otherwise arrive to the observer. This results in a diminishing of light even for a completely transparent planet, where no light is absorbed. We show that for planets occupying obits with large enough semimajor axis,
this refraction effect during transit results in an effective planetary radius, $R_p$, that is larger than that dictated by opacity.
Therefore, in some cases we would detect transits of completely transparent planets.

The structure of the paper is as follows. In \S\ref{assumptions} we lay out our model and simplifying assumptions. In \S\ref{spherical}
we derive the deflection angle through a spherically symmetric exponential atmosphere along with some analytic results for light trajectory
in such systems. Section \S\ref{images} discusses the enhancement of light outside transits and the decrease in light during transit. In \S\ref{Rayleigh}
we take the effects of scattering into account. We calculate the condition where scattering eliminates the additional light before and after
transits and where it  dominates the effective size of the planet during transits.

\section{Model assumptions}\label{assumptions}
We assume a planet in hydrostatic equilibrium with a spherically symmetric exponentially decaying density profile.
Since the deviation of the index of refraction of gas from unity $\delta n=n-1$ is proportional to its density, we also have an exponentially
decaying $\delta n$. The index of refraction $n$ as function of the distance from the center of the planet, $r$, is therefore:
\begin{equation}
\label{nr}
n(r)=1+\delta n(r)=1+\exp \left (  -\frac{r-R_1}{H}\right ),
\end{equation}
where $H$ is the scale height of the planetary atmosphere. Here we used the parameter $R_1$ as a nominal normalization of the index of refraction. It is the distance from the center where $\delta n$ is unity. In planets like Jupiter, it is roughly halfway to the center of the planet.
Therefore, $R_1$ is many scale heights inward of the usual definition of a planet's radius (e.g., where the optical depth radially outward is $\tau=2/3$).
The relation between $R_1$ and the radius of the planet will be clarified in \S \ref{ER} and in the summary.
The scale height $H$ is related to the planet's atmospheric temperature $T_p$ and its surface gravity $g$ by $H=\frac{k_{B}T_{p}}{\mu{g}}$,
where $k_B$ is the Boltzman constant, $T_p$ is the planet's atmospheric temperature, and $\mu$ is its mean molecular weight.
Note that we have used an exponential profile throughout the planet, but only the outer atmosphere matters.

We are mainly interested in light trajectories that pass a few scale heights near the radius of the planet, meaning in its atmosphere.
Inside the atmosphere, the light is diverted due to the changing index of refraction.
Our analytical results will be facilitated in the limit that the scale height of the planet is much smaller than the planet's radius which in turn is much smaller than the stellar radius $R_*$: $H \ll R_p \ll R_*$. Finally, we assume that the latter is much smaller than the semimajor axis of the planet $a$.
This last assumption implies that the angular deflection of the light, that bends over the planet's atmosphere and reaches a viewer at infinity  is
$\Delta \Theta \sim R_*/a\ll 1$.

\section{Light trajectories in a spherically symmetric medium}\label{spherical}

\subsection{Order of magnitude\label{oom}}

For small deflections, the innermost scale height towards which the light was directed dominates the deflection.
The distance traversed by an unbent ray within that scale height is $2\sqrt{R_pH}$. By definition of a scale height,
or equation (\ref{nr}), $\delta n$ changes significantly over a radial distance $H$.
Therefore, in a wide light ray, the part traveling at the top of the scale height  will be ahead of that at the bottom of the scale height
by a distance of order $(n-1)\sqrt{HR_p}$. We can therefore estimate the deflection angle to be $(n-1)\sqrt{R_p/H}$.
This calculation is valid for deflections which would shift the ray less than a single scale height over a distance of $\sqrt{R_pH}$. Therefore, for this result to hold,
the deflection angle must be less than $\delta \Theta \ll \sqrt{H/R_p}$.
Such small deflections can direct light towards the observer before transit only if $R_*/a \ll \sqrt{H/R_p}$.

\subsection{Fermat's Principle}

An elegant way of calculating the trajectory of light in the medium is by using Fermat's principle.
Following the spherical symmetry of the refraction index the trajectory takes place on a plane. Hence, we may use polar coordinates:
\begin{equation}
dt=\frac{n(r)}{c}\sqrt{dr^{2}+r^{2}d\theta ^{2}}
\end{equation}
and Fermat's principle reads
\begin{equation}
\delta \int dt= \delta \int  {\frac{n(r)}{c}\sqrt{1+r^{2}\left( d\theta \over dr\right)^{2}}dr}=0.
\end{equation}
Using the Euler-Lagrange equation, we obtain
\begin{equation}
{d \over dr} {n(r) r^2 {d\theta \over dr} \over \sqrt{1+r^{2}\left( d\theta \over dr\right)^{2}} }=0.
\end{equation}
or
\begin{equation}
\label{euler}
\frac{\partial \theta }{\partial r}=\mp \frac{b}{\sqrt{r^{4}n^2(r)-b^{2}r^{2}}}.
\end{equation}
Here $b$ is a constant of integration set by the initial conditions.
Far away from the planet, $n(r)_{r\to \infty }=1$, and the light travels on a straight line. Solving for $n(r)=1$ gives a straight line trajectory with a minimal radius $b$ and therefore $b$ is the impact parameter.
Taking the limit of equation (\ref{euler}) at large $r$ we obtain $d\theta/dr=b/r^2$, and therefore $b$ is the impact parameter
of the straight line that describes the trajectory far away from the planet, i.e. the impact parameter of the non deflected incoming light ray.

For a non-homogenous index of refraction, $(n\neq 1)$, the impact parameter $b$ and the minimum radius $r_{\min}$ do not coincide.
Since $d\theta/dr$ diverges at the minimum radius, we can find the minimum distance by setting the denominator in
equation (\ref{euler}) to zero. We obtain
\begin{equation}
\label{rmin}
r_{\min}n(r_{\min})=b
\end{equation}

An alternative way to describe the light trajectory, is to denote by $\alpha$ ($0\le \alpha \le 2\pi$) the angle between the propagation direction of the light ray and the radial direction. Then, $\tan \alpha=rd\theta/dr$.
Therefore, equation (\ref{euler}) can be elegantly written as
\begin{equation}
\label{nralpha}
rn(r)=b \sin \alpha .
\end{equation}
Equation (\ref{rmin}) is then a special case of equation (\ref{nralpha}) that describes the turning points, where the light ray is tangential so,
by definition, $\alpha=\pi/2$.
The functional form of $rn(r)$ therefore determines the properties of the light trajectories. Interestingly,
for our assumed density profile,
the function $rn(r)$ increases linearly steeply, reaching a maximum at about $\hat r \cong H$,
then decays roughly exponentially, reaching a minimum around $\tilde r  \cong R_1+H\ln R_1/H$, where $\tilde rn(\tilde r)  \cong \tilde r+H$.
Outward of that, it increases roughly linearly with $rn(r) \cong r$.
Note that we have assumed that $n-1$ grows exponentially all the way to the center.
Realistically, the density profile around the center grows more slowly.
Then, the maximum of $rn$ is not at $\hat r \cong H$, but much farther out.
The minimum we found at $\tilde r \cong R_1+H\ln R_1/H$ is up in the atmosphere,
where the exponential assumption is realistic, and therefore is correctly estimated.

We therefore conclude, that a light coming from infinity, with impact parameter $b$, will be deflected from the atmosphere of the planet,
if $b \gsim \tilde r+H$. In contrast,  for $b \lsim \tilde r+H$ light rays will deeply penetrate the planet and, may only be
reflected after arriving at $r \ll \hat r \cong H$. We note that all such rays will be absorbed in the planet, under any realistic opacity.
We now briefly mention two other solutions to equation (\ref{euler}) which are fully contained within the planet, and do not emerge out of it.

\subsubsection{Circular Trajectories}
At the minimum of $rn$ or at
\begin{equation}
r=\tilde r \cong R_1+H\ln R_1/H
\end{equation}
a light ray can have a circular orbit around the planet.
However, such orbits are unstable, and will ultimately either escape the atmosphere
or penetrate to large depth.

\subsubsection{Trapped Rays}
Since $rn$ may never increase above $b$ (as $\sin\alpha<1$)
we can have trapped rays that propagate between two radii, one close to the atmosphere, and one close to the center of the planet.
Such rays will be trapped between $r_1$ and $r_2$ satisfying:
\begin{equation}
\hat r < r_1\le r\le r_2 < \tilde r
\end{equation}
Again, these are likely to be absorbed for any realistic planet.

\subsection{The limit of small deflections}
We determine the angular deflection of light traveling through our spherically symmetric medium by integrating equation (\ref{euler}). For convenience, we subtract the result of a light traveling within a homogenous medium, $n=1$ with the same impact parameter $b$:
\begin{equation}
\Delta \Theta =2\int_{r_{\min}}^{\infty }(\frac{b/r}{\sqrt{r^{2}n^{2}(r)-b^2}} - \frac{b/r}{\sqrt{r^{2}-b^2}})dr
\end{equation}
Now, using a change of variables, $ x=\frac{r}{r_{\min}}$, and using (\ref{rmin}) we get

\begin{equation}
\Delta \Theta =2\int_{1}^{\infty }(\frac{1/x}{\sqrt{x^{2}\frac{n(x)^{2}}{n(r_{\min})^{2}}-1}} - \frac{1/x}{\sqrt{x^{2}-1}})dx
\end{equation}
Under the assumption that $r_{\min}/H \gg 1$ yet $\delta n r_{\min}/H \ll 1$ we can expand $\frac{n(x)^{2}}{n(r_{\min})^{2}}$
to the first order in $\delta n$ and get an expression for the angular shift:
\begin{equation}
\Delta \Theta = \delta n(b)\sqrt{2\pi\frac{b}{H}}
\label{dtheta1}
\end{equation}
This is the basic result we will use next. It is equivalent to what we found using order of magnitude derivation in \S\ref{oom},
but includes an additional factor of $\sqrt{2\pi}$.

\section{Lensed Stellar Images}\label{images}

Equipped with an analytic expression for the angular deflection, we determine below the star's images to a viewer at infinity.
Using these images we then estimate the change in the observed luminosity due to atmospheric lensing.

A light arriving from any point on the star could pass by both sides of the planet and be deflected towards the observer at infinity.
For example in figure \ref{schematic}, point A has two images $A'$ and $A''$.
If the center of the planet is in between the source and the image as projected on the sky we refer to the deflection as deflection through 
the far side of the planet. Otherwise, if, in projection, the source and the image are both on the same side of the center of the planet we
refer to the deflection as deflection through the near side of the planet. 
Out of transit, the deflection through the far side of the planet would yield a secondary image of the star that appears within the planet's atmosphere and has a crescent shape.
The light that is deflected though the near side of the planet and reaches an observer at infinity, is generally way up in planet's atmosphere and is hardly deflected given the exponential nature of the index of refraction (for example, points $A''$ and $C''$ in figure \ref{schematic}). These rays produce the simple, almost unperturbed, disk-like image of the star away from transit.
The secondary, crescent like, image changes the ordinary light curve away from, but close to, transits. In order to estimate the change in the observed flux we find the area of the crescent image as refraction preserves the specific intensity.

\subsection{Evaluation of secondary image size}
We denote the star radius as  $R_{\star }$, the semi-major axis of the planetary orbit as $a$ and the projected distance between the star and the planet on the major-axis as $XR_{\star }$.

\begin{figure}[h!]
  \centering
  \includegraphics[width=0.65\textwidth, viewport=0 200 770 500, clip]{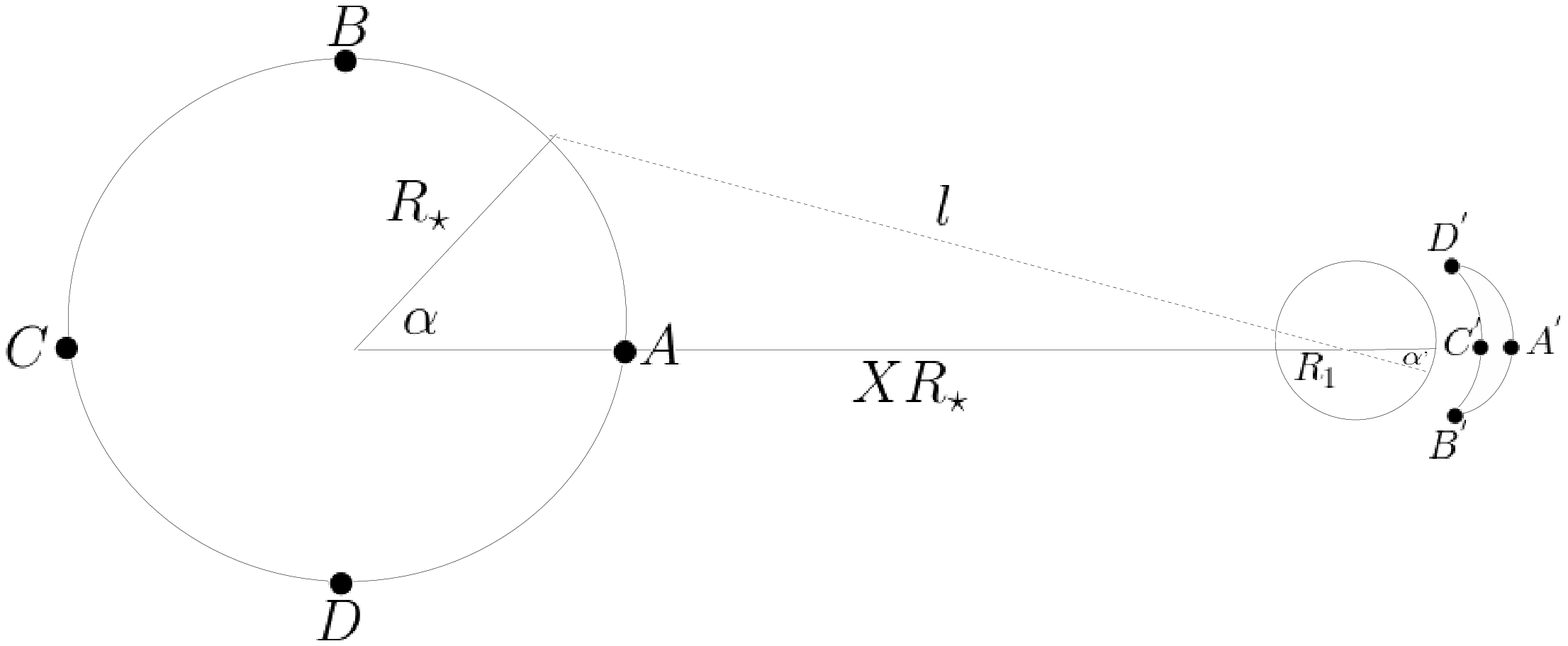}
  \includegraphics[width=0.65\textwidth]{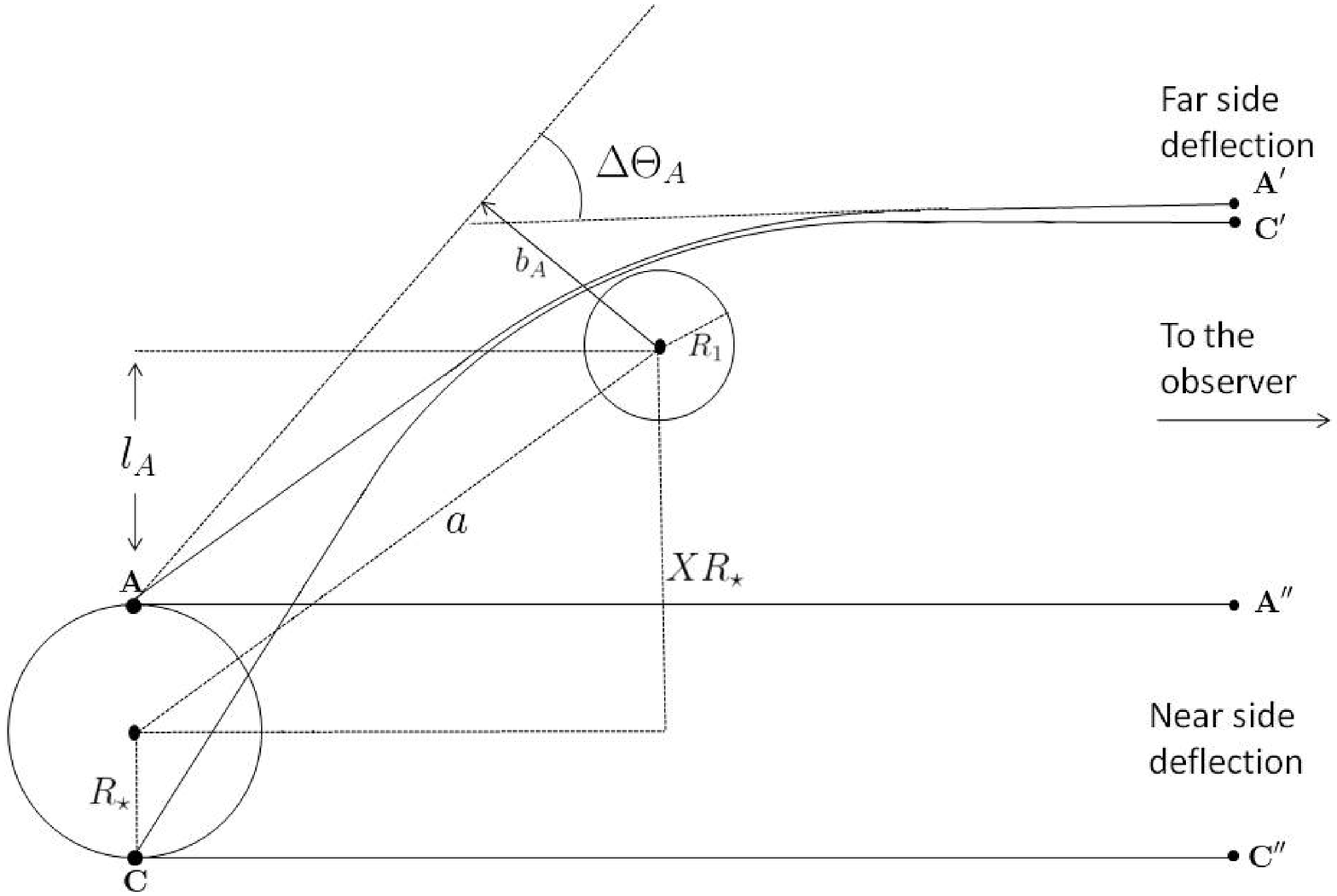}
  \caption{A schematic diagram showing the crescent shape of the secondary stellar image.
 The upper panel shows the projection on the sky while the lower panel shows the plane that contains the observer and the center
 of the planet and star, allowing to follow ray trajectories.
   The points $A$, $B$, $C$ and $D$ are mapped by the atmospheric lensing onto points $A'$ $B'$ $C'$ and $D'$. On the lower
   panel, points $A$ and $C$ are also shown to arrive practically undeflected to points $A''$ and $C''$.
  An arbitrary source point in the star located at an angle $\alpha$, is mapped into $\alpha'$ in the atmosphere of the planet.
  \label{schematic}}
\end{figure}
Now we must formulate a proper way of integrating over the star's edge which will be mapped, due to refraction through the far side of the planet's atmosphere, to the crescent shaped image. Using Green's theorem for a closed curve integral we can calculate the crescent area. If we denote the crescent's area and the distance of each point on the crescent from the planet's center by $S'$ and $b$ respectively, then
\begin{equation}
 S'=\int^{\alpha'_{\max}}_{\alpha'_{\min}} \frac{1}{2}b^2d\alpha'.
 \label{sp}
\end{equation}
To evaluate the integral, we change integration variable from $\alpha'$ to $\alpha$ using
\begin{equation}
\label{alphaalphap}
\tan{\alpha'}=\frac{\sin{\alpha}}{X-\cos{\alpha}}.
\end{equation}
For the light to arrive at the observer, the deflection angle must satisfy
\begin{equation}
\sin \Delta \Theta={l + b \over a}
\label{dtheta2}
\end{equation}
where from Figure \ref{schematic}, we see that
\begin{equation}
l=(X-\cos \alpha)R_*/\cos \alpha'.
\end{equation}
We can therefore obtain $b(\alpha)$ by equating equations (\ref{dtheta1}) and (\ref{dtheta2}). While in  \S\ref{num} we numerically solve these equations for $b$, analytical expressions are given in \S\ref{awayor} and \S\ref{nearor} under some limits.

\subsection{The Effective Radius of The Planet \label{ER} }

We have used the parameter $R_1$ to normalize our density or index of refraction profiles. During the transit itself, light is deflected away from the observer.
We can now calculate the effective radius of the planet as the radius of the dark spot in the star's image caused by this deflection.
In the middle of a central transit, i.e., when the center of the planet covers the center of the star as projected on the sky, the deflection angle that determines the effective radius is given by  $\Delta \Theta_{CT} =(R_{\star }+R_p)/a$.
This is the deflection of light coming from the edge of the star and refracted in the far side of the planet in order to reach the distant observer. Using this deflection angle we obtain
\begin{equation}
\label{Rp}
R_p \equiv b\left( \Delta \Theta_{CT} \right) \cong R_{1}+\frac{H}{2}\ln\left ( \frac{2\pi R_pa^2}{HR_{\star}^2} \right ).
\end{equation}
This is an implicit equation since $R_p$ appears on the right hand side as well. However, since it only appears in the logarithm,
any value for the radius can be used there with negligible errors. Exact solutions can be found iteratively with ease.
Note, that the above is not a standard definition of the planet radius as it depends explicitly on its distance from the star.
Yet, our $R_p$ as defined above is quite similar to the planet's radius defined in other ways. Comparison
between our effective radius for occultations $R_p$ with other
definitions of the radius of the planet related to atmospheric absorption or scattering are given in \S\ref{Rayleigh}.
This definition of the planet radius, allows for a simpler equation of the deflection angle at small deflections:
\begin{equation}
\label{dtheta3}
\Delta \Theta={R_* \over a}\exp \left( R_p-b \over H \right)
\end{equation}

\subsection{Images by numerical calculation\label{num}}
\begin{figure}[h!]
  \centering
             \includegraphics[width=0.5\textwidth,viewport=55 350 555 530,clip]{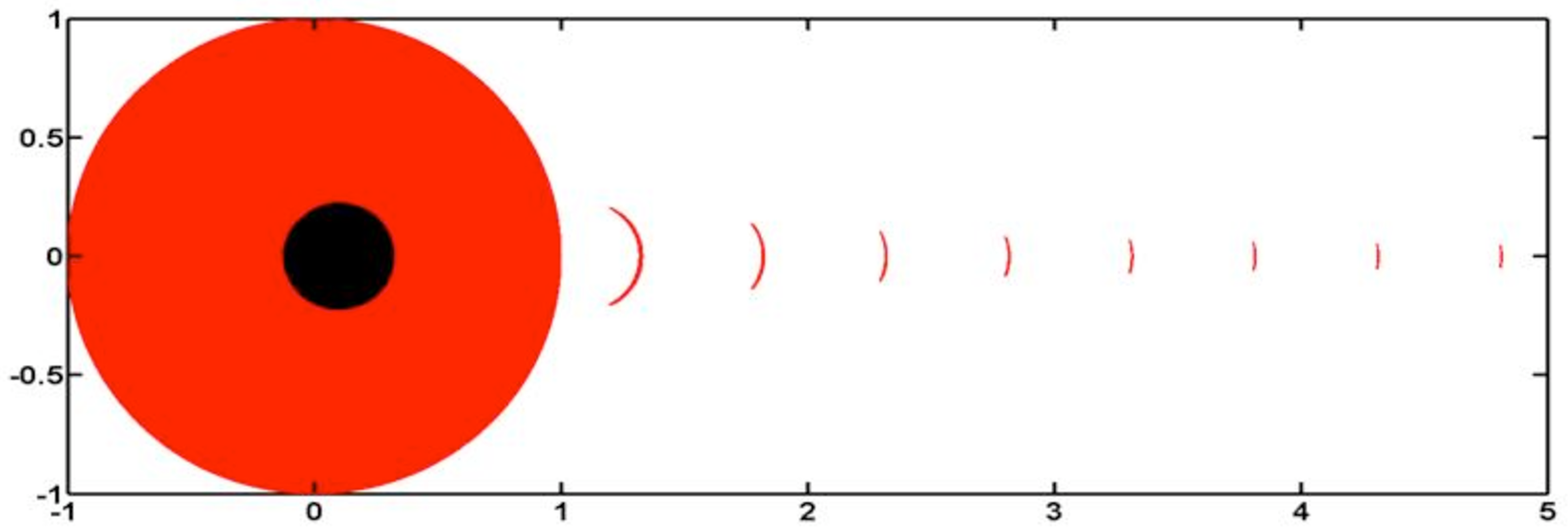}
  \caption{
  \label{cressun}
  The image of the star as refracted through an exponential planetary atmosphere plotted for different values of planet star separation, $X$.
  The axis are measures in units of star radius. For $X>1$, an additional crescent shape image is seen through the planet's atmosphere.
  When $X<1$ the star is being eclipsed even if the planet is completely transparent.}
\end{figure}

\begin{figure}[h!]
  \centering
      \includegraphics[width=0.5\textwidth,viewport=30 300 585 550,clip]{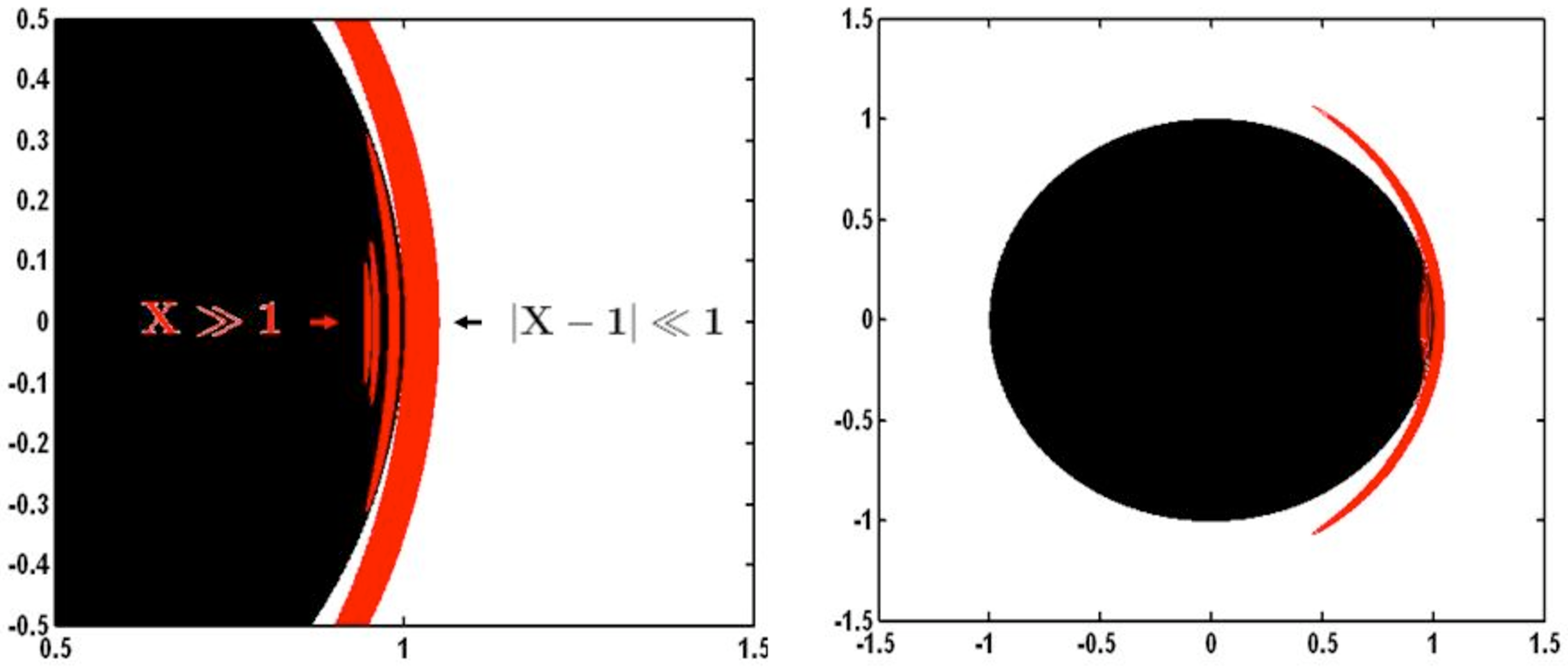}
  \caption{
  \label{cresplanet}
The crescent as it appears in the planetary atmosphere for different values of the separation parameter $X$. In this figure the origin is
fixed to the planet center, and the coordinates are in units of the planet's radius $R_p$.
As $X$ decreases the crescent moves away from the planet, becomes wider and covers a bigger angle. }
\end{figure}

We numerically integrate equation (\ref{sp}) and plot the images of the star as seen by an observer at infinity.
In order that the effect will be easily noticed we chose the following (not realistic) parameters
$a=100R_{\star}$, $R_{\star}=8R_{1}$ and $R_{1}=40H$.
For these parameters, equation (\ref{Rp}) then implies $R_p=1.7R_1$.
The image shapes are plotted in Figures (\ref{cressun}) and (\ref{cresplanet}).
We can see that as $X$ diminishes the secondary image becomes larger.

For $X<1$ the planet occults the star and an eclipse will be observed.
To understand that, we take as an example the case where the center of the planet and the center of the star coincide in projection on the sky.
In this case, every point on the limb of the star will either be almost unperturbed when it is deflected from the near side of the planet,
while deflections from the far side will pass at a distance of $R_p$ from the center of the planet (per the definition of $R_p$).
Therefore, a dark spot will be observed in the center of the star, with radius $R_p$.
The change in observed flux affected by the eclipse is given by the ratio of the areas $S'/S=R_p^2/R_{\star}^2$.
Symmetry, of course, dictates a bright spot at the center of the dark one since a light ray that would come from the center of the
star passing through the center of the planet will arrive without deflection to the observer. However, as we have discussed below equation
(\ref{nralpha}) such rays penetrate so deep into the planet and are always absorbed. We therefore ignore them.

\begin{figure}[h!]
  \centering
      \includegraphics[width=0.55\textwidth, viewport=50 180 820 500,clip]{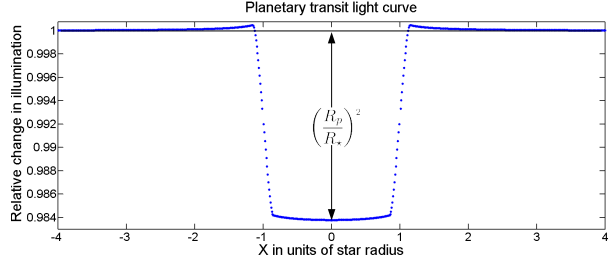}
  \caption{The light curve of a star during a completely transparent planetary transit
  while considering atmospheric lensing, for $a=100R_*$, $R_*=8R_p$ and  $R_p=100H$. Just before and just after the transit,
  the stellar light is increased due to refraction in the atmosphere of the planet. Analytical expressions for these shoulders
  are given in equations (\ref{spfar}), (\ref{spfarnum}) \& (\ref{spnear}).
  The curved bottom of the transit appears without taking limb darkening into account. Similar curvature in transit was obtained
  by \cite{HuS02}.  
  \label{lc1}}
\end{figure}

This decrease of $(R_p/R_*)^2$ for transparent planets is similar to the one of opaque planets.
Transparent planets produce similar transits to opaque ones!
The light curve that correspond to the parameters mentioned earlier is shown in Figure (\ref{lc1}).
The crescent shaped secondary image of the star does not yield an analytical expression for a generic of $X$.
An analytical expression for the crescent's size can be obtained under several limits. First we study the case where the planet is far away from occultation regime, which corresponds to $X\gg 1$, and then we study the crescent size just before transit, where $R_p/R_*<X-1\ll 1$.

\subsection{Away from occultation regime\label{awayor}}

From Figure (\ref{schematic}), we obtain
\begin{equation}
\Delta \Theta = \frac{R_{\star}}{a}\sqrt{X^2-2X\cos{\alpha}+1}\cong \frac{XR_{\star}}{a}\left ( 1-\frac{\cos{\alpha}}{X} \right )
\end{equation}
We also have an expression for $\Delta \Theta $ from (\ref{dtheta3}).
Since $b\gg H$ and both $R_{p}$ and $b$ differ by a few $H$'s we solve for $b(\alpha)$ as
\begin{equation}
b\cong R_{p}-H\ln\left ( X-\cos \alpha \right ) \cong R_{p}-H\ln X+{H \over X} \cos \alpha
\end{equation}
Noting that
for  $\left ( X\gg 1 \right )$ equation (\ref{alphaalphap}) can be approximated as
\begin{equation}
d\alpha'=\frac{\cos{\alpha}}{X}d\alpha
\end{equation}
We can estimate the secondary image size
\begin{equation}
S'=\oint {b^2 \over 2} d\alpha'=\frac{\pi HR_{p}}{X^2}
\end{equation}
Now if we assume that the light is not scattered or absorbed, and ignore limb darkening of the star,
the brightness will be uniform and identical over both the direct and lensed stellar images and
the fractional increase in light arriving to the observer due to atmospheric lensing is
\begin{equation}
\label{spfar}
\frac{S'}{S}=\frac{HR_{p}}{ X^2 R_{\star}^2}.
\end{equation}
or, substituting the expression for $H$,
\begin{eqnarray}
\label{spfarnum}
\frac{S'}{S} & = & 6.4\times 10^{-6} \left(a \over {\rm 1AU} \right)^{-1/2} \left( R_p \over R_J \right)^3 \cr
& \times & \left( M_p \over M_J \right)^{-1}  \left( L \over L_\odot \right)^{1/4} \left( R_* \over R_\odot \right)^{-2} X^{-2}
\end{eqnarray}

\subsection{Near occultation regime\label{nearor}}
\begin{figure}[h!]
  \centering
      \includegraphics[width=0.55\textwidth,viewport=30 200 600 600,clip]{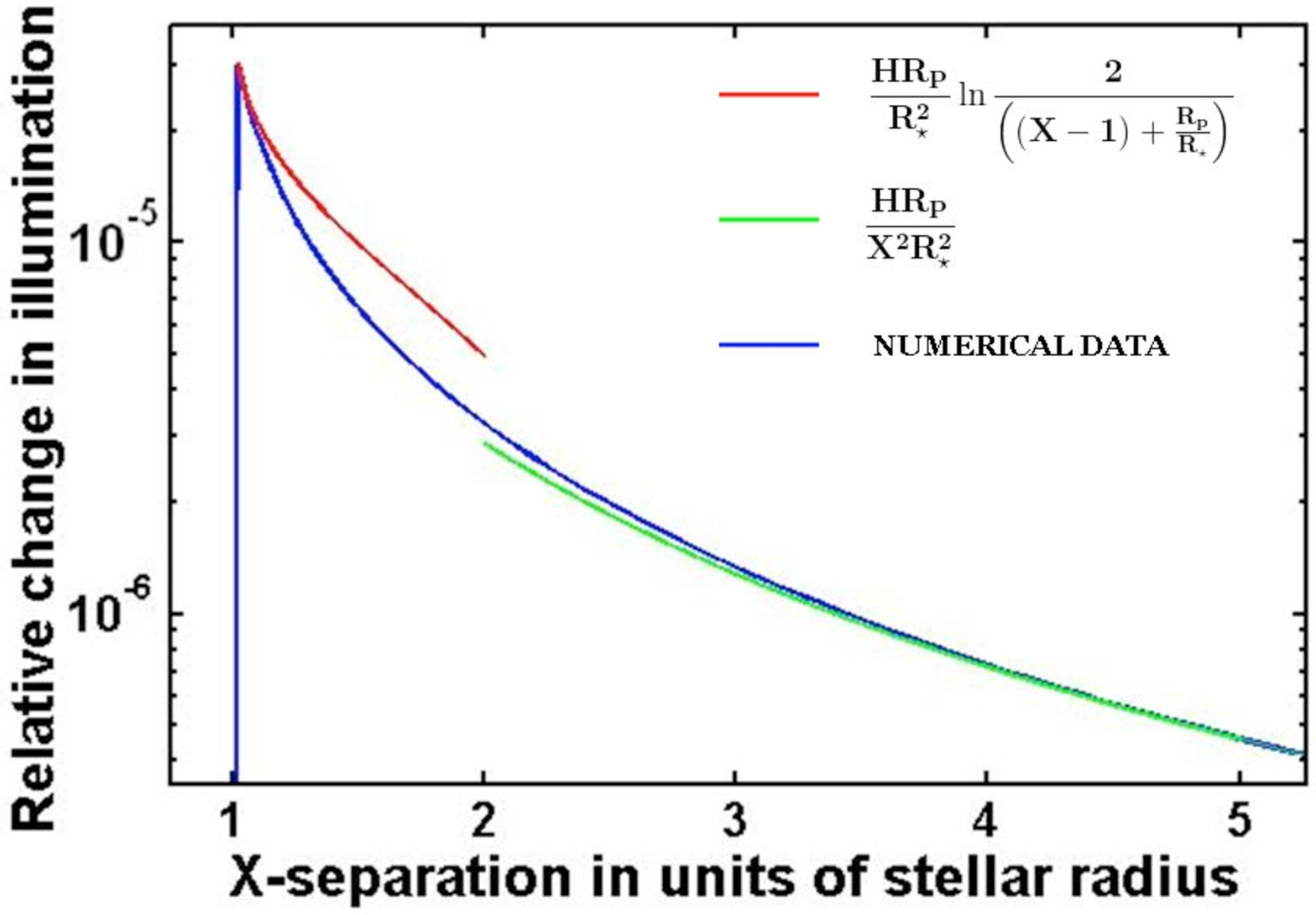}
  \caption{The shoulders in transit light curve, ignoring extinction.
 The numerical curve is plotted as blue line for $a=100R_{\star}$, $R_{\star}=10R_{p}$ and $R_{p}=100H$. \label{ill}
For comparison, our analytical estimates far away from transits (green line, equation (\ref{spfar})) and very close to transit (red line, equation (\ref{spnear})) are also shown for the same parameters. It can be seen that these provide excellent approximations within their range of validity.}
\end{figure}

Observing the numerical plot of the planetary transit light curve shown at Figure (\ref{lc1}),
and the crescent images plotted in Figure (\ref{cressun}) and (\ref{cresplanet}),
one can easily notice that the effect is most substantial when $X$ is near unity.
Hence, we derive now asymptotic expressions in the limit of $ \frac{R_p}{R_{\star }} <X-1 \ll 1$.
In this limit the crescent radial width does not change significantly until its edges at $\alpha'$ of almost  $\pm \frac{\pi}{2}$.
The area of the crescent is therefore $\pi HR_{p}\Delta b$ where $\Delta b$ is it's central width. To estimate $\Delta b$,
we use equation (\ref{dtheta3}):
\begin{equation}
b \cong R_{p}- H\ln\left ( \Delta \Theta  \frac{a}{R_{\star}} \right )
\end{equation}
Following the notation of Figure (\ref{schematic}), we get
\begin{equation}
 \Delta b\cong b_{A'}-b_{C'}=H\ln \left ( \frac{\Delta \Theta _{A'}}{\Delta \Theta _{C'}} \right )
\end{equation}
Noting that
\begin{equation}
\label{ApCp}
 \Delta \Theta _{A'}=\frac{\left ( X+1 \right )R_{\star }+R_{p}}{a}  \;\;\;\; \Delta \Theta _{C'}=\frac{\left ( X-1 \right )R_{\star }+R_{p}}{a}
\end{equation}
together with $ \frac{R_p}{R_{\star }} <X-1 \ll 1$ and $R_{\star} \gg R_{p}$ we finally get
\begin{equation}
 \Delta b\cong H\ln \left ( \frac{2}{ X-1 + \frac{R_{p}}{R_{\star}}} \right ).
\end{equation}
The crescenst's area is then
\begin{equation}
 S'\cong \pi R_{p}H\ln \left ( \frac{2}{ X-1 + \frac{R_{p}}{R_{\star}}} \right ),
\end{equation}
so that the fractional increase in observed flux is given by
\begin{equation}
\label{spnear}
 \frac{S'}{S}\cong  \frac{R_{p}H}{R_{\star}^2}\ln \left ( \frac{2}{ X-1 + \frac{R_{p}}{R_{\star}}} \right ).
\end{equation}
This is up to a factor of $  \ln\frac{R_{\star}}{R_{p}}\approx 2.3$ larger than what the extrapolation of equation (20) to $X=1$ would imply.
Figure (\ref{ill}) shows how the asymptotic estimates (\ref{spfar}) and (\ref{spnear}) coincide with the numerical calculation.

\subsection{Conservation of light?}

We have shown so far that light is added to the observer when the planet is out of transit.
There, the effect is of order $HR_p/R_*^2$, and when averaged\footnote{This average is not over an orbit, but over all orbits with all inclinations}
over all possible positions of the planet around the star we have $HR_p/a^2$.
However, during transit, refraction causes a decrease of order $(R_p/R_*)^2$.  Averaged over all possible planet positions we have $R_p^2/a^2$.
The two effects, therefore, do not cancel.
To settle this apparent discrepancy, one must take into account the light that penetrated deeply into the planet.
The amount of that light is $R_p^2/a^2$, but it is being
spread roughly equally over $4\pi$ steradians as viewed from the planet.
Therefore, for a non realistic, truly transparent planet, the planet will shine
out of transit with brightness smaller than that of the star by $(R_p/a)^2$.

The effect of the crescent that we calculated, which result in shoulders around transits, is therefore not offsetting the light refracted
away during transit. The latter is offset mostly by deeply penetrating rays, that would allow that planet to shine anywhere in the orbit.
The shoulders provide only a small correction to the above. Together, averaged over all possible planet positions, the shoulders
and the deeply penetrating rays which show up away from transit, cancel the decrease of light during transit exactly.
However, in any realistic situation deeply penetrating rays will be absorbed, and therefore less light, on average, will arrive to the observer.

\section{Extinction by Rayleigh scattering}\label{Rayleigh}
In order to determine the relevance of refraction one must estimate the starlight extinction as it passes through the planet's atmosphere.
Rayleigh scattering is a minimal cause of extinction.
We ignore molecular absorption and clouds. We also ignore limb darkening, and denote the uniform, unextincted, stellar surface brightness by $I_0$.
The observed brightness after a given path through the planet's atmosphere is $I=I_{0}\exp \left ( -\tau  \right )$,
where $\tau$ is the optical depth:
\begin{equation}
\tau=\int \sigma N dl.
\end{equation}
Here, $\sigma=\frac{32\pi^3}{3\lambda^4}\frac{( n-1 )^2}{N^2}$ is the total cross section for Rayleigh scattering per molecule of the gas, $n$ is the index of refraction, $N$ is the number of molecules per unit volume,$\lambda$ is the wave length.
The integration takes part over the path of light.
Here we assume that the planet has a Hydrogen atmosphere, in this case we have
\begin{equation}
d \equiv  \frac{ n-1}{N}= 3.046 {\rm \frac{cm^3}{mol}}
\end{equation}
For an atmosphere in hydrostatic equilibrium the gas density decays exponentially over the scale height $H \ll R_p$, so that
the principal contribution to the optical depth comes from the integration over a few scale heights near the impact parameter $b\cong R_p$.
Therefore, for a light ray with negligible deflection,
\begin{equation}
\tau=\int_{-\infty}^\infty  \sigma N_{max}\exp \left ( -\frac{\sqrt{l^2+b^2}-b^2}{H} \right )dl,
\end{equation}
where $l$ is the coordinate along the light trajectory measured from the point of closest approach to the planet center.
By expanding the root around small $l/b$, and assuming $b \gg H$, we obtain an analytic expression for the optical depth:
\begin{equation}
\tau =\frac{\sigma }{d}H\Delta \Theta = \frac{32\pi^3d}{3\lambda^4} H \Delta \Theta
\end{equation}

Now, using the expression for the cross section $\sigma$ and assuming that the planet's temperature is determined from thermal equilibrium
with albedo $A$,
\begin{equation}
\label{taugen}
\tau=\frac{32\pi^{3}dk_BT_{\star}R_p^2}{3\sqrt{2}\lambda^{4}\mu GM_{p}}\left ( \frac{R_{\star}}{a} \right )^{1/2}\left ( 1-A \right )^{1/4}\Delta\Theta.
\end{equation}

We now derive asymptotic results far away from transit, $X \gg1$, as well as just before transits $R_p/R_*<X-1\ll 1$.

\subsection{Away from occultation regime\label{exfar}}
When $X \gg 1$, the refraction angle of the close and distant edges of the star do not vary considerably, $\Delta \Theta = \frac{XR_{\star}}{a}$.
Therefore, from equation (\ref{taugen}), the optical depth is uniform. The crescent specific intensity is constant,
attenuated simply by $e^{-\tau}$, where
\begin{eqnarray}
\label{taufar}
\tau & =& 0.067\left ( \frac{\lambda _{\rm red}}{\lambda} \right )^4 \left ( \frac{L_{\star}}{L_{\odot}} \right )^{1/4}\left ( \frac{R_{p}}{R_{\j}} \right )^{2} \nonumber \\
 & \times & \left ( \frac{M_{p}}{M_{\j}} \right )\left ( \frac{R_{\star}}{R_{\odot}} \right )\left ( \frac{1AU}{a} \right )^{3/2}\left ( 1-A \right )^{1/4}X.
\end{eqnarray}
Here we have used red light with $\lambda_{\rm red}=650$~nm as our fiducial wavelength. Clearly, the optical depth in
the blue part of the spectrum is significantly larger.

The fractional flux from the crescent before extinction, is given by equation (\ref{spfar}), it decreases with the planet's
semimajor axis $a$ as $a^{-1/2}$. On the other hand $\tau$ decreases as $a^{-3/2}$. The transmitted light is therefore maximal when
\begin{equation}
\tau = 1/3.
\end{equation}
We therefore define
\begin{eqnarray}
\label{a13}
a_{1/3} & = &0.34~{\rm AU}~\left ( \frac{\lambda _{\rm red}}{\lambda} \right )^{8/3} \left ( \frac{L_{\star}}{L_{\odot}} \right )^{1/6}\left ( \frac{R_{p}}{R_{\j}} \right )^{4/3} \nonumber \\
& \times & \left ( \frac{M_{p}}{M_{\j}} \right )^{2/3}\left ( \frac{R_{\star}}{R_{\odot}} \right )^{2/3}\left ( 1-A \right )^{1/6}
\end{eqnarray}
to be the semimajor axis where optical depth is $1/3$ for deflections at angles $R_*/a$.
For our fiducial parameters, $a_{1/3}$ corresponding to an orbital period of 74 days.

For $a \gg a_{1/3}$ the refraction effect will be practically unextincted, with magnitude given by equation (\ref{spfar}), for any $X<X_{\max}$ given by
\begin{equation}
X_{max} \cong (a/a_{1/3})^{2/3}.
\end{equation}
 For $X>X_{\max}$ the crescent flux will decay exponentially with $X$ due to Rayleigh scattering.

 \subsection{Near occultation regime\label{exnear}}
When the planet center as projected on the sky is less than a stellar radius away from the star's edge (point A in figure \ref{schematic}), the refraction
angles of the close and distant edges of the star (points A and C on figure \ref{schematic}) differ considerably. The light from point A is deflected much less than $R_*/a$ and therefore will be significantly less extincted than we estimate in \S\ref{exfar}. Hence, in order to estimate the attenuation of the star light that is refracted through the planet atmosphere, one needs to integrate along the crescent. In that case the fractional size of the effect, taking extinction into account, is given by
\begin{equation}
 \left . \frac{S'}{S} \right |_{\rm ext}=\frac{S'}{S} \frac{\int_{R_{\rm in}}^{R_{\rm out}} e^{ -\tau\left ( r \right)}  dr }{\int_{R_{\rm in}}^{R_{\rm out}}dr} = \frac{R_{p}}{R_{\star}^2} \int_{R_{\rm in}}^{R_{\rm out}} e^{ -\tau\left ( r \right) }dr,
\end{equation}
where $R_{\rm out}$ and $R_{\rm in}$ the maximal and minimal radius of the crescent as measured from the center of the planet, respectively.
The denominator integral is simply the crescent radial width in the $ \frac{R_p}{R_{\star }} <X-1 \ll 1$ regime.
Changing variables from $r$ to $\tau$ we obtain
\begin{equation}
\left . \frac{S'}{S} \right |_{\rm ext} = \frac{\int_{R_{\rm in}}^{R_{\rm out}} e^{ -\tau\left ( r \right)} dr }{\int_{R_{\rm in}}^{R_{\rm out}}dr} = \frac{R_{p}H}{R_{\star}^2} \int_{\tau_{\rm out}}^{\tau_{\rm in}}d \tau \frac{e ^{-\tau}}{\tau}
\end{equation}
Here $\tau_{\rm in}$ and $\tau_{\rm out}$ refer to the optical depth associated with the refracted light that sets the inner and outer edges of the crescent. Explicitly
\begin{equation}
\tau_{\rm in}=\frac{\sigma }{d}H\Delta \Theta _{A'} \;\;\;   \tau_{\rm out}=\frac{\sigma }{d}H\Delta \Theta _{C'}
\end{equation}
From equation (\ref{ApCp}) we see that $\tau_{\rm in}$ is given by equation (\ref{taufar}) with $X=2$ while $\tau_{\rm out}$ is obtained if $X$ is
replaced by $X-1+R_p/R_*$. We examine the attenuation scaling in a few different regimes.

\begin{itemize}
\item
Both $\tau_{\rm in}, \tau_{\rm out} \gg 1$.  In this regime, the crescent luminosity will undergo considerable extinction:
\begin{equation}
\frac{S'}{S}_{\rm ext} \cong \frac{HR_{p}}{R_{\star}^2}\frac{e^{-\tau_{\rm out}}}{\tau_{\rm out}}
\end{equation}
This applies for $a \ll (2R_p/3R_*)^{2/3}a_{1/3}$ or, for our fiducial parameters $a \ll  0.06$~AU, i.e., a period shorter than 5 days. Most current
transits, therefore, fall in this category, making refraction effects difficult to see. The additional light in this case grows exponentially in time
as the planet approaches transit. Most light arrives when the planet is about $R_p$ away from transit.
Therefore, the duration of the increased light would be shorter than the duration of the transit by $R_p/2R_*$, i.e. about a factor of 20.
For hot Jupiters, transits, which last about 3 hours, should be accompanied by slightly bright edges for about 10 minutes.

\item
The crescent is only partially extincted. $\tau_{\rm in} \gg 1$ but $\tau_{\rm out} \ll 1$.
Here the crescent luminosity changes considerably along its radial width,
so the crescent luminosity is dominated by regions where $\tau<1$.
We obtain:
\begin{equation}
    \frac{S'}{S}_{\rm ext} \cong \frac{HR_{p}}{R_{\star}^2} \ln \left ( \frac{1}{\tau_{\rm out}} \right ),
\end{equation}
relevant for $3^{-2/3}a_{1/3} \gg a \gg (2R_p/3R_*)^{2/3} a_{1/3}$ or for our fiducial parameters
$0.16{\rm ~AU} > a > 0.06$~AU or periods between 5 and 25 days.
The fractional flux undergoes two competing effects. On one hand the area of the crescent
decreases as $a^{-1/2}$, on the other hand $\tau_{\rm out} \propto a^{-3/2}$ decreases with $a$, allowing a larger fraction of the light to pass through.
The maximum is obtained where $\tau_{\rm out}=e^{-3}\cong 0.05$. However, at such large semimajor axis, given our fiducial parameter,
$\tau_{\rm in} = (R_*/R_p)\tau_{\rm out} \cong 0.5 <1$.
Therefore, the maximum fractional luminosity of the crescent is obtained at the transition between this and the next regime, when $\tau_{\rm in}=1$,
with orbital periods of 25 days.
There,
\begin{equation}
{S' \over S} \cong {H R_p \over R_*^2} \ln (R_*/R_p) \cong 3.6 \times 10^{-5}.
\end{equation}
This is the largest effect refraction can produce. The numerical value can be slightly higher if the planet's radius is larger
than that of Jupiter and if its mass is somewhat lower, allowing for a larger scale height at a given temperature.

\item
Both $\tau_{\rm in}, \tau_{\rm out} \ll 1$. Here the crescent luminosity is hardly attenuated, so that its
fractional luminosity is given by (\ref{spnear}). It decreases with semimajor axis showing again, that the maximum is obtained around
 $\tau_{\rm in}=1$.

\end{itemize}

 \section{Summary \label{summ} }

 We have derived the possible paths of light passing through a planet with exponential deviation of the index of refraction from unity. Though we explore exotic trapped rays and rays following circular orbits, we focus on small deflections, as given by our equation (\ref{dtheta1}).

We explore the size and shape of the image of the star as refracted through the planet's atmosphere. The size of the image
is given by equation (\ref{spfar}) far away from transit while the largest and least extinct effect should be seen just before transit, and its
magnitude is given by equation (\ref{spnear}). For planets with short orbital periods, refraction involves relatively large deflections, and will therefore
be reduced by extinction. On the other hand, planets with large orbital period have cold surfaces resulting in a small effect.
The most noticeable effect would
be for planets with orbital period of about $25$ days.
For extrasolar planets with $R_p \cong 1.5R_J$, like HD~209458 where $R_p\cong 1.4 R_J$ and $M \cong 0.7M_J$ \citep{BCG+01}, the fractional flux increase would be as large as $10^{-4}$, if it was located on a 25 days orbit.  This is readily observable with current instruments.

\cite{HuS02} have also considered refraction by planetary atmospheres. However, they focused on caustics, which are important
for point sources. These are relevant for refraction of background stars by atmospheres of planets in our solar system.
We instead, were interested in the case where the source is very large, smearing out the effects of caustics. We have also ignored the effect of oblateness that they considered.

For HD~209458 on its observed 3.5~days orbit, the refraction effect would be extincted with optical depth of about $\tau_{\rm out}=1.7$. The refractive effect, even given this extinction would be about $10^{-5}$. It would last for about 10 minutes just before and just after transit. This is still detectable in existing or future data.  If the enhancement of light is detected, it could provide a direct measurement of the scale height  of extrasolar planet.

The timescale for the brightest part of the refractive effect is somewhat shorter than the duration of the transit, especially if extinction is involved
for planets with short orbital periods.

We also show, that during the eclipse, refraction of stellar light away from the observer may cause a signal similar to that of a regular geometric eclipse. We define the radius $R_p$ as the effective radius of the planet as observed by a refractive eclipse. Indeed, for
any planet on orbital periods longer than 25 days, $R_p$, the radius where refraction effect control the transit occur higher up in the atmosphere
that $R_{ray}$, where Rayleigh scattering is important. Using equation (\ref{a13}) we find that these two radii are related by
\begin{equation}
R_p=R_{ray}+{3 \over 2} H\ln\left( 2^{2/3} a \over a_{1/3} \right)
\end{equation}
Of course, other kinds of extinction may be in place, most notably clouds, which we have ignored. Those may increase the effective
extinction radius above $R_{ray}$ and perhaps beyond our effective radius $R_p$.

Finally, we mention that refraction effects would have only a slight modification on the Rossiter - McLaughlin effect, but perhaps a large effect
on the absorbed spectrum during transits. If $R_p-R_{\rm ray} \gg H$ such that refraction rather than extinction determine the eclipse, less absorption would be observed in the spectrum.

\acknowledgments We thank Oded Aharonson and Peter Goldreich for helpful discussions.
RS is partially supported by an ERC \& IRG grants and a Packard Fellowship.

\end{document}